\documentclass[default]{aastex701}

\def\sgra{SgrA$^\ast$}
\def\mpc{\mathrm{mpc}}
\def\pc{\mathrm{pc}}
\def\myr{\mathrm{Myr}}

\usepackage{subfigure}
\usepackage{amsmath}

\begin{document}

\title{Binary Dynamics and Mergers in a Stellar Disk and the Formation of Dusty Objects in the Galactic Center}

\author[orcid=0000-0002-8793-3246, gname=Myank, sname=Singhal]{Myank Singhal}

\affiliation{Astronomical Institute, Faculty of Mathematics and Physics,V. Hole\v{s}ovi\v{c}k\'{a}ch 2,Prague,18000, Czech Republic}
\email[show]{myank.singhal@matfyz.cuni.cz}  

\author[orcid=0000-0003-1924-8834]{Ladislav \v{S}ubr} 
\affiliation{Astronomical Institute, Faculty of Mathematics and Physics,V. Hole\v{s}ovi\v{c}k\'{a}ch 2,Prague,18000, Czech Republic}
\email{subr@sirrah.troja.mff.cuni.cz}

\author[orcid=0000-0001-8594-3689]{Jaroslav Haas}
\affiliation{Astronomical Institute, Faculty of Mathematics and Physics,V. Hole\v{s}ovi\v{c}k\'{a}ch 2,Prague,18000, Czech Republic}
\email{haas@sirrah.troja.mff.cuni.cz}

\author[orcid=0000-0002-9850-2708, gname=Florian, sname=Peißker]{Florian Pei{\ss}ker}
\affiliation{I.Physikalisches Institut, Universit\"at zu K\"oln, Z\"ulpicher Str. 77,K\"oln, 50937, Germany}
\email{peissker@ph1.uni-koeln.de}

\author[orcid=0000-0001-6450-1187]{Michal Zaja\v{c}ek}
\affiliation{Department of Theoretical Physics and Astrophysics, Faculty of Science, Masaryk University, Kotlá\v{r}ská 2, Brno, 61137, Czech Republic}
\email{zajacek@mail.muni.cz}

\begin{abstract}
For more than a decade, observers have been tracking a population of enigmatic dusty objects orbiting the central supermassive black hole of the Milky Way, Sagittarius A$^\star$ (\sgra). In one interpretation, these objects are products of stellar mergers enshrouded in gaseous-dusty envelopes, which account for their infrared excess, emission lines, and stability against the tidal forces of \sgra. Recent near-infrared observations of D9, one of the dusty objects, revealed that it embeds a spectroscopic stellar binary. In this work, we present full-fledged N-body simulations of an eccentric disk of binary stars similar to the young stellar disk observed around \sgra. We find that some of the binaries in the disk are naturally drawn towards \sgra and merge there due to Kozai-Lidov (KL) oscillations in their orbits. Together with the observed nature of D9, these findings represent a self-consistent picture of how the population of the observed dusty objects could have been brought into existence. The binary fraction in our N-body model evolves to increase with distance from \sgra, which is also in agreement with the observational constraints. Our simulations also suggest that beside the dusty objects, stellar mergers are present within the young stellar disk, which could be identified as chemically peculiar or fast-rotating stars. Our results show that dynamical evolution of binary stars in the young stellar disk plays a crucial role in shaping the stellar content in the Galactic Center. 
\end{abstract}



\section{Introduction}

The center of our Galaxy harbors a supermassive black hole (SMBH) associated with the compact and variable radio source \sgra with a mass of about $4.3\times10^6 M_{\odot}$ \citep{gravity_collaboration_deep_2022,EHT2022}; see also \citet{Eckart2017}, \citet{Genzel2022}, and \citet{ciurlo_sagittarius_2025} for recent reviews. It is 
embedded in an old (nuclear) star cluster whose total mass is about six 
times higher than that of the SMBH and a characteristic (half-light) radius of roughly 4 pc \citep{schodel_surface_2014}. The presence of such an old and 
massive cluster around \sgra has always been expected \citep{young_numerical_1980,genzel_stellar_2003,merritt_dynamics_2006}, but
what came as a surprise was the discovery of very young stars closer
than 0.5 pc from \sgra \citep{eckartObservationsStellarProper1996,ghezHighProperMotionStars1998,aliKinematicStructureGalactic2020,Ciurlo2020,Peissker2024a,peisskerClosingGapFollowup2025}. Moreover, it turned out that a significant
fraction \citep[20-50\%;][]{yelda_properties_2014} of these
stars are orbiting \sgra coherently in multiple well-defined disc-like structure, with the most prominent one spanning between $40\ \mathrm{mpc} \lesssim r \lesssim 400\ \mathrm{mpc} $ and known as the clockwise disk \citep{levin_stellar_2003,paumard_two_2006,vonfellenbergYoungStarsGalactic2022b}. Quite in contrast, closer than 40 mpc from \sgra 
(within the so-called S-cluster), the orbits of the young stars are more or less isotropically oriented \citep{ghez_first_2003,eisenhauer_sinfoni_2005}; however, see \citet{aliKinematicStructureGalactic2020} who suggests that the young stars form various disk like structures as well. The origin of the young stars in the 
innermost half-parsec of our Galaxy and the configuration of their 
orbits around \sgra is still debated.

Recent observations of the immediate surroundings of {\sgra} have 
revealed an even more mysterious population of dust-embedded objects \citep{gillessenGasCloudIts2012,phifer_keck_2013,Eckart2013,Valencia2015,Ciurlo2020,Peissker2020b, Peissker2024a,peisskerClosingGapFollowup2025}. They are located 
mainly in the S-cluster, and their orbits around \sgra \ are similar to 
those of the young stars. The nature and origin of these dusty objects is puzzling, with two prominent ideas discussed in the literature. One possibility is that these are pure gas clouds in orbit around \sgra \citep{gillessenGasCloudIts2012,Gillessen2025}. Alternatively, other authors suggest that the dusty envelope contains a stellar core.
Within this view, there are two main possibilities: in one, these dusty objects contain a forming star \citep{murray-clay_disruption_2012,scoville_galactic_2013,zajacek_dust-enshrouded_2014}. Interestingly, this hypothesis immediately raises another puzzling question: How can individual stars form so close to \sgra \citep{ghez_first_2003,morrisMassiveStarFormation1993}?

Another interpretation of the dusty objects is that they could result from stellar binary mergers which eject material that condenses into dusty envelopes. Notably, these various formation scenarios are not mutually exclusive. The recent discovery of a stellar binary behind the dusty curtain of the object D9 provides an example of stellar cores in dusty objects, while also demonstrating that dusty envelopes can develop in binary systems before merging \citep{peisskerBinarySystemCluster2024}. 

A necessary prerequisite for the stellar binary merger formation scenario is a
source of stellar binaries. A natural source for these stellar binaries can be the clockwise disk \citep{chuEvidenceDecreasedBinary2023,gautam_estimate_2024}.
\citet{subrPropertiesHypervelocityStars2016} studied the evolution of binary systems made up of point masses in a disk-like system, where the binaries undergo KL oscillations due to the disk. KL oscillations lead to an increase in eccentricity of the outer orbits (the orbits of the binary system's center of mass around \sgra), resulting in close passages near {\sgra} and the production of hyper-velocity stars (HVS) and S-stars through the Hills mechanism \citep{hillsHypervelocityTidalStars1988}. At the same time, the inner orbit (orbit of the binary system around their own center of mass) of the binaries can undergo KL oscillations due to the influence of \sgra. The KL oscillations increase the eccentricity of the inner orbit, resulting in close passages of the two stars that can drive the individual binaries towards mergers \citep{naoz_eccentric_2016,stephan_merging_2016,stephan_fate_2019, Ciurlo2020}. 

Until now, however, a self-consistent model for the transport of binaries from their possible birthplaces in the clockwise disk to the S-cluster—with mergers occurring ``on the run"—has been missing. While previous studies have explored the dynamical fate of binaries near an isotropic Galactic Center, they often operate under different parameter spaces or environmental assumptions \citep{bradnickStellarBinariesGalactic2017,mcguireSemiAnalyticalModelEvolution2026}. For instance, \citet{bradnickStellarBinariesGalactic2017} focused primarily on tight binaries with separations up to $0.3\ \mathrm{au}$. In contrast, our work considers a broader range of $0.1$–$100\ \mathrm{au}$. This distinction is critical: the wider separation range significantly alters the ratio of stars that merge versus those that separate.


In this work, we use direct N-body modeling to show that the evolution of a disk of mutually interacting binary stars—driven by the eccentric Kozai-Lidov (KL) mechanism within its own potential—naturally transports binaries into the S-cluster. The mutual interaction results in a different evolution from that seen in other recent works studying the dynamical evolution of binary systems in isolation for the production of S-stars through the Hills mechanism \citep{bradnickStellarBinariesGalactic2017,sersanteDynamicsRecapturesEjections2025}. The production of S-stars via the Hills mechanism within this framework is also subject to ongoing simulations and will be addressed in a subsequent paper, as in our current simulations, stars are modeled as rigid spheres without tidal effects. These tidal effects can dissipate energy, resulting in orbital decay and circularization.

We show in our work that during close passages to \sgra, the stellar binary can undergo chaotic interactions that change its orbital elements. Although we currently cannot quantify how tidal dissipation will affect the results, it presents a compelling avenue for future study. 

Here, we focus on the spatial distribution and the binary fraction as a function of distance from \sgra. We compare these merger products with observed dust-embedded objects in the Galactic Center, demonstrating that this transport and merger channel is a promising explanation for the population currently observed.

The structure of this paper is as follows. In Section~\ref{sec:setup} we introduce the model setup including initial conditions, numerical methods, the treatment of stellar mergers, and information about KL oscillations, relativistic damping, and tidal effects. In Section~\ref{sec:results} we present the main results concerning the stellar-merger temporal, mass, and orbital distributions, including their comparison with the observed dusty objects. We conclude with Section~\ref{sec:conclusion}. 

\section{Model \& Methods} \label{sec:setup}

\subsection{Setup}

We simulate the evolution of a disk of binary stars in the Galactic Center in orbit around \sgra. The setup consists of a disk of 2000 stars orbiting a massive central SMBH, mimicking \sgra, of mass $ M_{\bullet} =4 \times 10^6 M_\odot$ \citep{ghez_first_2003,eisenhauer_sinfoni_2005,gillessen_monitoring_2009,gillessen_orbit_2009,yeldaIncreasingScientificReturn2010}. The stellar masses for disk stars are sampled from a power-law distribution function $n_m \propto m^{-1.5}$ and $1 M_\odot < m < 150 M_\odot$ motivated by the analysis of young stars in the Galactic Center and star formation in molecular clouds \citep{luLARPOPULATIONSCENTRAL2013,zhang_probing_2024}. The total mass of the disk is $M_{\mathrm{disk}} \approx 24500 M_\odot$.

In the disk, all stars are paired to form binaries with a mass ratio close to unity, motivated by the observed binaries in the near vicinity of the SMBH, GCIRS 16SW and  IRS 16NE having mass ratio close to unity \citep{martins_gcirs_2006,pfuhl_massive_2014}. This is in accordance with \citet{kobulnickyNewLookBinary2007} who find that massive stars tend to form binaries with similar masses. For simplicity, we have also assembled the low-mass binaries to be of similar masses \footnote{Since the binary pairs are not exact twins, we do not exclude the effects of the eccentric KL mechanism, which can go to zero for systems with the same masses \citep{naoz_eccentric_2016}.}.  The binaries are in circular orbits (the ``inner" orbits henceforth), and their orbital angular momentum is parallel to the disk's angular momentum. The semi-major axis of the inner orbits ($a_\mathrm{inner}$) is sampled from the Öpik distribution function, $n_{a_\mathrm{inner}} \propto a_\mathrm{inner}^{-1}$\citep{kobulnickyNewLookBinary2007,lingDistributionSeparationsDMSA2004}. The range of $a_\mathrm{inner}$ is $0.1-100$ au, which gives us a mixture of tight and wide binaries, with a preference for tighter ones. Also, the binaries are paired so that they do not merge at the start of the simulation. 
    
These binaries are placed in the disk with their center of masses orbiting the SMBH (the ``outer" orbits henceforth). The semi-major axes of the outer orbits ($a_\mathrm{outer}$) follow a power-law distribution $n_{a_\mathrm{outer}} \propto a_\mathrm{outer}^{-1}$. The disk's inner edge is at $a_\mathrm{min}=0.02$pc while its outer edge is at $a_\mathrm{max}=0.4$pc. The eccentricity of the outer orbits follows a radial gradient according to 
$e=0.9 (a_\mathrm{outer}-a_{min})/(a_{max}-a_{min})$, 
motivated by simulations that model the formation of the young stellar disk observed in the Galactic Center by an infalling molecular gas cloud in the vicinity of an SMBH \citep{bonnellStarFormationSupermassive2008,mapelliSituFormationSgrA2012,generozov_forming_2022}. The binaries are also verified to be stable against the tidal potential of \sgra at the start of the simulation.  The eccentricity vectors of the orbits are parallel, which means that they are mutually aligned and are not randomly oriented. Hence, binaries form a disk which provides a non-axisymmetric perturbation to the gravitational potential of \sgra. These binaries are placed on orbits with an inclination that follows the distribution function $n_i \propto \sin{i}$ with $i \in [ 0,2.5^\circ ]$. The impact of the binary separation ($a_\mathrm{inner}$) and the inner edge of the disk ($a_\mathrm{min}$) are not critical, and the overall physics remains the same, which is discussed in \autoref{app:ini_impact}.


To save computational resources, we omit the nuclear stellar cluster. Previous work shows its effect on KL damping is negligible unless the cluster mass significantly exceeds 
$\sim 4 M_\mathrm{disk}$ with the cluster enclosed within a radius of $a_\mathrm{max}$\citep{subrPropertiesHypervelocityStars2016}. The stellar mass estimated to be enclosed in the galactic center up to 1 pc away from \sgra is $0.8 - 1.2 \times10^6 M_\odot$ based on various different normalization factors \citep{schodel_distribution_2018}. This could mean that the mass enclosed within $a_\mathrm{max}$ can be up to an order of magnitude higher than $M_\mathrm{disk}$ depending on how the mass is distributed. However, as shown in \citet{subrPropertiesHypervelocityStars2016}, it impacts binaries with pericenter distances below $1 \mpc$ around \sgra, affecting the number of binaries that could undergo the Hills mechanism, which is outside the scope of this work. Other limitations of the numerical setup are discussed in \autoref{app:limitations}. 
\subsection{Integrator}

We perform NBODY integrations using NBODY6 \citep{Aarseth_2003}. This code integrates the equations of motion for each body within the mutual interaction of other bodies. We use a modified version of the code which increases logging frequency and uses a weighted decision criterion (weighted by mass of the bodies) for adding stars to the neighbor list \citep{subrPropertiesHypervelocityStars2016}.

We allow stellar collisions throughout the integration of the equations of motion. The stars merge into a single object if their separation gets smaller than the collisional radius, 

\begin{equation}
    R_\textrm{coll}=1.7 \left( \frac{m_1+m_2}{2 m_1} \right)^{1/3} r_1^*,
    \label{eq:rcoll}
\end{equation} 
where, $m_1$ and
$m_2$ are the stellar masses and $r^*_{1}$ is the larger of the stellar radii \citep[Eq. 9.21;][]{Aarseth_2003}. We further assume that the merger product retains all of the mass of the parent stars. This is done for simplicity, as the goal of the work is to understand the dynamical evolution of binaries.

Using this integrator, we run 200 independent simulations with initial conditions sampled from the parameters described above. However, some of the simulations could not be used due to large accumulated errors during their integration. So we use a reduced sample of 182 simulations, which is sufficient to characterize the system’s general evolution.  

\subsection{Kozai-Lidov oscillations}
Due to the overall geometry of the investigated system, one of the driving dynamical processes is the KL oscillations of the individual stellar (or binary star) orbits around \sgra induced by the gravitational influence of the disk itself. In addition to this, however, the inner orbits of the components of individual binary stars will undergo such oscillations due to the gravitational pull of \sgra.

In their simplest (quadrupole) mode, KL oscillations occur on the characteristic KL timescale, $T_\mathrm{KL}$. The quadrupole approximation corresponds to the configuration in which the orbit of the subject particle around the dominant body (\sgra or binary center of mass in our case) is perturbed by a distant perturber whose orbit around the dominant body is circular. The KL timescale is given by \citep{Kozai1962, Lidov1962}:
\begin{equation}
    T_\mathrm{KL} \equiv \frac{M_\mathrm{cent}}{M_\mathrm{pert}} \frac{a_\mathrm{pert}^3}{a \sqrt{G M_\mathrm{cent} a}} (1-e^2_\mathrm{pert})^{3/2} ,
    \label{KL_timescale}
\end{equation}
where ${a_\mathrm{pert}}$ denotes the semi-major axis of the orbit of the perturber,${e_\mathrm{pert}}$ is the eccentricity of the perturber, $a$ is the semi-major axis of the subject particle (both quantities are measured with respect to the dominant body), $M_\mathrm{cent}$ and $M_\mathrm{pert}$ stand for the mass of the dominant body and the perturber, respectively, and $G$ is the gravitational constant. If the orbit of the perturber around the dominant body is eccentric, the quadrupole KL oscillations are typically modulated on a longer timescale and octupole order effects need to be included (see \citet{naoz_eccentric_2016} for further information). 

The parameters given above in our model can be used to estimate characteristic timescales of the KL oscillations, $T_\mathrm{KL}$ (~\autoref{KL_timescale}). In the case of the KL oscillations of the outer binary in our model, the dominant body is represented by \sgra, $M_\mathrm{cent}=M_\bullet$, and the disk serves as the perturber $M_\mathrm{pert}=M_\mathrm{disk}$, as well as the reservoir of the subject orbits (mostly from its inner parts). For $a_\mathrm{pert}\approx 0.12 \mathrm{pc}$ and $e_\mathrm{pert}\approx 0.45$ (roughly corresponding to the center of mass of the stellar disk),   and $a \approx 20 \mathrm{mpc}$ (initial inner edge of the disk), $T_\mathrm{KL}$ is about $0.52$ Myrs, which decreases for $a\approx 40 \mpc$ to about $0.18$ Myrs. This KL timescale calculated here is just an approximate one for the disk as a single perturber, while the realistic KL timescale due to the disk configuration cannot be estimated here.  
The KL timescale of the inner binary, $T_\mathrm{KL}$, in our model depends sensitively on the distance of the binary from \sgra and can therefore vary by orders of magnitude between systems. As an illustrative example, consider a binary with component masses of $1.5\,M_\odot$ and $0.5\,M_\odot$, an inner semi-major axis $a_\mathrm{inner}=1\,\mathrm{au}$, and initially circular inner and outer orbits. If binary orbits \sgra with $a_\mathrm{outer}=20\,\mpc$, the inner KL time scale is $\sim 0.004\,\myr$, corresponding to a ratio $T_\mathrm{KL,outer}/T_\mathrm{KL,inner}\sim 134$. For a wider orbit with $a_\mathrm{KL,outer}=90\,\mpc$, this ratio changes dramatically to $T_\mathrm{KL,outer}/T_\mathrm{KL,inner}\sim 0.15$. This strong variation demonstrates that the hierarchy of timescales can change substantially as the outer orbit evolves due to KL oscillations, implying that the inner KL evolution may proceed in a non-adiabatic regime. In particular, during phases of high outer eccentricity, $T_\mathrm{KL}$ for the inner binary can decrease rapidly, and the ratio of $T_\mathrm{KL,outer}/T_\mathrm{KL,inner}$ can evolve as the outer orbit evolves further.






We are omitting relativistic effects due to \sgra in our simulations to reduce computational cost. This removes damping of KL oscillations due to the relativistic precession of orbits. KL oscillations arise from coherent secular torques exerted by a distant perturber on an inclined inner orbit, in the double-averaged approximation, these torques vary slowly and enable a resonant exchange between eccentricity and inclination. However, when relativistic apsidal precession becomes sufficiently rapid, the argument of periapsis circulates on a time scale shorter than the KL timescale. This rapid precession disrupts the phase coherence required for eccentricity growth, thereby suppressing KL oscillations.

\citet{karasEnhancedActivityMassive2007a} studied how relativistic precession and an extended nuclear cluster dampens KL oscillations for stars with small $s$ or high $e$. They showed a reduction in stars with extremely close flybys ($\sim d < 10^{-2} \mpc$ for \sgra) to the central massive black hole, which should not affect the results of our study. We also present results in \autoref{sec:gr_consequences} that show that while relativistic corrections of the outer orbits do modify KL dynamics, they do not compromise the results presented in this work.

Studies have investigated the effects of short-range forces on KL dynamics, including relativistic precession and tidal forces, which can significantly affect inner KL oscillations \citep{wuPlanetMigrationBinary2003, fabryckyShrinkingBinaryPlanetary2007, liuSuppressionExtremeOrbital2015, naoz_eccentric_2016}. In \autoref{sec:gr_consequences}, we present more detailed results on how inner KL oscillations are affected by relativistic precession. We find that for a binary whose inner KL oscillations are damped due to relativistic precession, the evolution of the outer orbit can cause the binary to evolve into regions where the KL oscillations are stronger, which can cause the binaries to merge.

Apart from relativistic precession, tidal dissipation within stellar binaries modifies the KL oscillations. In hierarchical triples, tidal dissipation can convert Kozai-Lidov (KL) cycles of high-eccentricity into orbital shrinkage rather than prompt mergers \citep{fabryckyShrinkingBinaryPlanetary2007,antoniniSecularEvolutionCompact2012,naozMergersObliquitiesStellar2014,prodanSECULAREVOLUTIONBINARIES2015,stephan_merging_2016,stephan_fate_2019}. Unlike standard setups where the outer orbit evolves slowly, our binaries orbit \sgra with the stellar disk driving rapid KL oscillations on the outer orbit. This creates non-adiabatic variations in the inner secular evolution, triggering rapid eccentricity excitation during close passages to \sgra (Section \ref{sec:inner_orbit_gr}). As in our simulations, stars are currently modeled as rigid spheres; tidal effects are omitted. Standard equilibrium tides typically operate on megayear timescales \citep{hutTidalEvolutionClose1981,fragioneMassiveBinaryStar2019}. Recent work suggests chaotic tides on highly eccentric orbits can circularize binaries within a few orbits—particularly for initial inner separations of $\sim 1\ \mathrm{au}$ compared to $0.1\ \mathrm{au}$ \citep{dodiciDynamicalEvolutionStellar2025,hao-tsehuangFormationCloseBinaries2025}. However, while those models rely on outer-orbit evolution via gigayear-scale relaxation, our systems undergo rapid dynamical interactions on much shorter timescales. As a result, we cannot predict how these chaotic tides will affect our simulations, and they provide a compelling avenue for future work. We also would like to add that chaotic interactions with \sgra or other disk stars, which we demonstrate later in Section \ref{sec:results}, can change the orbital elements of the inner binary such that relativistic precession no longer suppresses inner-KL oscillations \citep{winter-granicBinaryMergersCenters2024}. Although tidal dissipation may save a subset of high-eccentricity binaries from merging, it is unlikely to eliminate the merger population entirely, making its precise impact an important subject for future study.

\section{Results and discussion} \label{sec:results}

\subsection{Stellar Mergers}
\label{sec:stellar_mergers}

\begin{figure}
    \centering
    \includegraphics[width=\linewidth]{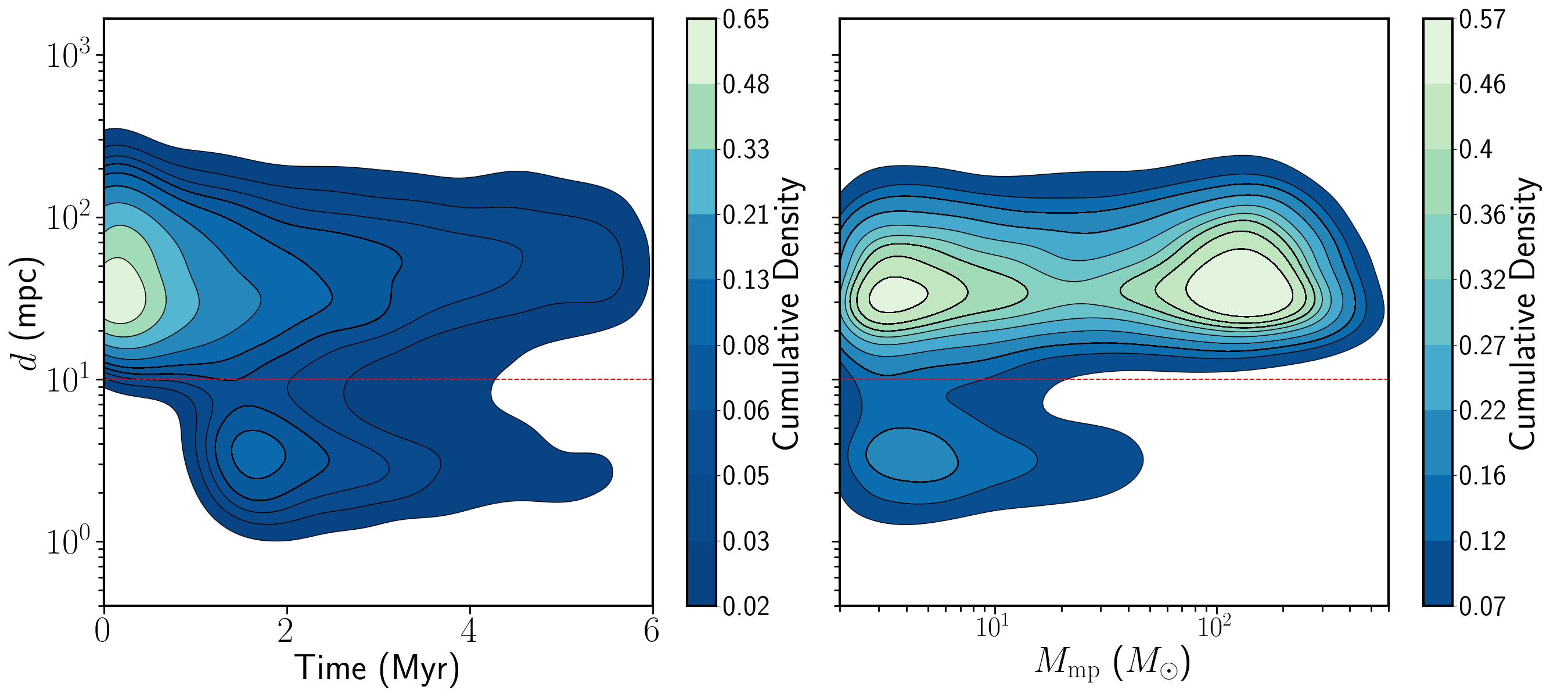}
    \caption{Properties of stellar mergers. Left and right panels show the cumulative density distribution of radial distance from {\sgra} ($d$) at which collisions occurred as a function of time and mass of the merger product, $M_{\mathrm{mp}}$, respectively. The colours represent the area outside of which a certain percentage of the population resides.The red dotted line at $d=10$ mpc is an arbitrarily selected dividing line based on the properties of the mergers, where the region of $d>10$ mpc is the disk body and the region of $d\leq 10$ mpc is the innermost region of the Galactic Center, within the S cluster.}
    \label{fig:merged_binaries_raduis_mass_time}
\end{figure}

In the simulations, we first examine the locations and other properties of the mergers. \autoref{fig:merged_binaries_raduis_mass_time} illustrates the radial distance, $d$, from \sgra at which each binary merged, along with the time and mass of the resulting product, $M_\mathrm{mp}$. Two distinct populations emerge: most coalescences occur within the main body of the parent stellar disk, at radial distances $d\gtrsim10\,\mpc$ from {\sgra}. Those taking place well below the inner edge of the stellar disk form a separate group. They start to appear at $T\approx1\,\myr$ after the beginning of integration with the peak of their production at $T\approx2\,\myr$. 
This delay is comparable to the time required for eccentric (octupole) KL oscillations to increase the outer orbits eccentricity due to the parent disk, while also undergoing stochastic interactions with other stars in the disk that change both the inner and outer orbital elements \citep{haas_rich_2016,stephan_merging_2016,subrPropertiesHypervelocityStars2016,stephan_fate_2019}. 

We now discuss the properties of the two groups of mergers distinguished by $d$ at the time of merging. 

\subsubsection{Collisions in the main body of the stellar disc ($d\gtrsim10\mpc$)}

\begin{figure}
  \centering

    \subfigure[]{\includegraphics[width=0.49\linewidth]{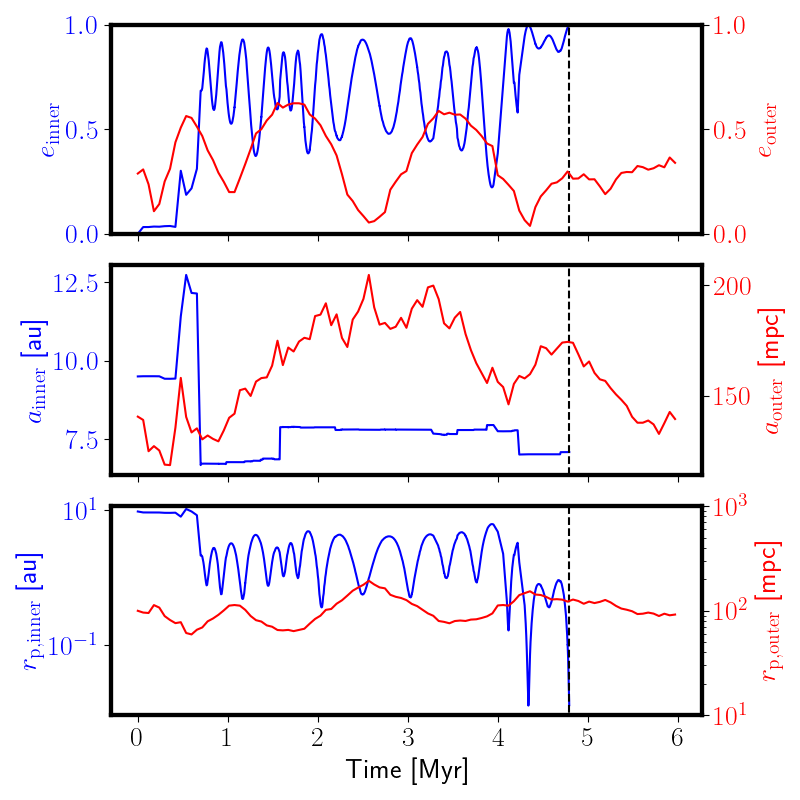}} 
  \subfigure[]{\includegraphics[width=0.49\linewidth]{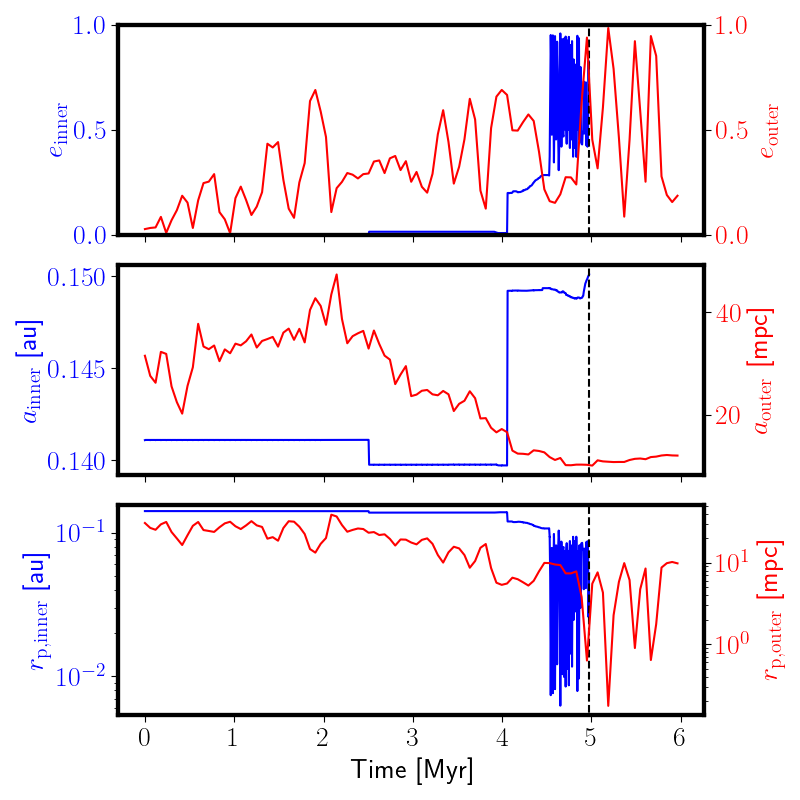}}

  \caption{Example of the orbital evolution of two different binary systems that merge. The blue lines show the orbital parameters of the inner orbit with the scale given on the left, and the red lines represent the outer orbital parameters and the scale is given on the right. The top panel shows the eccentricity evolution, middle panel represents the semi-major axis, and the bottom panel depicts the temporal evolution of the pericenter distance. The merger time is depicted with vertical dashed lines. In panel (a), the inner binary undergoes eccentricity oscillations before the merger occurs at $T=4.79\ \myr, d=225.2\ \mpc$, while in panel (b) we see a binary merging due to rapid eccentricity oscillations induced by evolution of the outer orbit around \sgra at $T=4.97\ \myr, d=4.40\ \mpc$.  }
 
\label{fig:merger_examples}
\end{figure}

Binaries within the disk begin to merge shortly after the integration starts since they mutually perturb each other in a relatively dense environment. These perturbations during close encounters can result in sudden and large changes of the inner orbital elements. This results in binary disruptions and mergers, whose rate decays in time in accordance with slow disintegration of the stellar disk, which leads to a decrease in stellar density. Beside that, there are mergers that occur due to the gradual growth of inner eccentricity—most likely driven by KL oscillations due to \sgra—ultimately leading to stellar collision. In \autoref{fig:merger_examples} (a), we show an example of a merger that takes place at $d=225.2\mpc$ due to oscillations of eccentricity. The same example also shows instantaneous changes of semi-major axis caused by close interactions with other stars at various time points, with some example at $T\approx0.7 \myr$ and $T\approx4.3 \myr$.

Mergers within the disk involve stars from the whole mass range with two prominent peaks in mass distribution of the merger products (right panel of \autoref{fig:merged_binaries_raduis_mass_time}). We attribute this shape of the distribution function to the fact that low-mass stars are more abundant in the system, hence, the number of their mergers is also higher. The high-mass peak of the distribution function comes from the fact that the most massive stars have larger radii, i.e. they are closer to the merger limit (\autoref{eq:rcoll}). As the initial setup just verifies that the binaries were not below the merger limit before the start of the simulation, even small perturbations in the orbital elements can cause a large number of high mass binaries to merge at the start of the simulation.

\subsubsection{Collisions below the inner edge of disk ($d<10\mpc$)}
The less prominent group of mergers that occur below the inner edge of the stellar disk are driven towards the coalescence by a similar binary evolution. In some aspects, however, the evolution is quantitatively different because now the binary undergoes evolution of the outer orbit, likely due to KL oscillations induced by the disk \citep{haas_rich_2016,subrPropertiesHypervelocityStars2016}. When the outer orbit enters a phase with a very large eccentricity, the binary suffers from close encounters with {\sgra} in contrast to stellar-mass perturbers in the disk.

During the orbit around \sgra, the inner binary is subject to perturbations from \sgra, which can result in KL oscillations or chaotic changes in orbital elements. This can lead to a merger that retains the outer orbital properties of the parent binary around \sgra. In \autoref{fig:merger_examples}, panel (b), we show a merger during a close pericenter passage near \sgra ($d=4.4\mpc$), marked by rapid oscillation in the inner eccentricity. If we assume that these eccentricity oscillations are caused by KL oscillations, the proximity to {\sgra} shortens $T_\mathrm{KL}$ for the inner binary, resulting in rapid KL oscillations that can trigger a prompt merger.

During close pericenter passages, the binary gets perturbed by {\sgra}. This could result in the separation of the two stars with one of them being accelerated to a high velocity, leaving the other on an orbit more tightly bound to {\sgra}, than was the progenitor binary \citep[the Hills mechanism;][]{hillsHypervelocityTidalStars1988,antonini_tidal_2011,subrPropertiesHypervelocityStars2016}. Our simulations indicate that mergers are roughly three times less likely than separations occurring near \sgra. Strong perturbation to the inner orbit may also result in chaotic interactions, in which the binary may become regular again once it travels farther from \sgra, but with different inner orbital elements. 

The mergers that occur below the inner edge of the disk are preferably from the low-mass end of the mass spectrum (right panel of \autoref{fig:merged_binaries_raduis_mass_time}). This is mainly a selection effect -- most of the massive binaries that were susceptible to collisions due to minor perturbations from other stars have merged already within $\lesssim 0.5\,\myr$ in the main body of the stellar disk. Only the more abundant lower-mass binaries have survived in a sufficiently large amount to later times when KL oscillations push a (relatively small) subset of them to highly eccentric orbits.

\subsection{Orbital elements of Merger products}
\label{sec:orb_ele_merger_products}

\begin{figure}
    \centering
    \includegraphics[width=\linewidth]{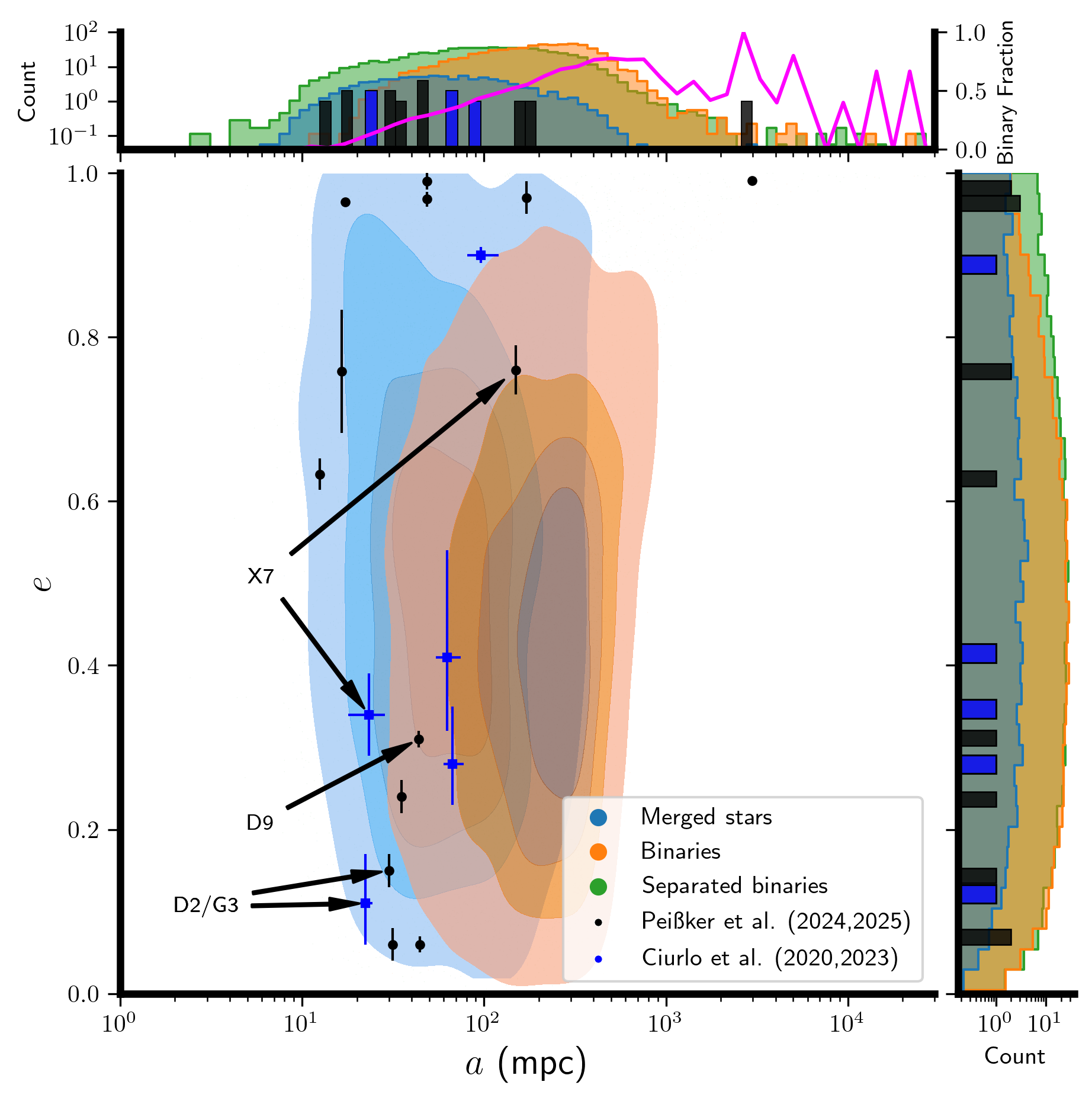}
    \caption{In the central density plot, we showcase the final eccentricity and semi-major axis of observed and simulated objects of total mass less than $20 M_\odot$. The density plot shows the surviving binaries and the merged stars, with each increasing colour bin corresponding to an additional 20\% population outside the colour boundary. The blue dots in the figure show the orbital elements of dust enshrouded objects as given by \citet{Ciurlo2020,ciurlo_swansong_2023} and the black dots are orbital elements for dust enshrouded objects given by \citet{Peissker2024a,peisskerBinarySystemCluster2024,peisskerClosingGapFollowup2025}. Out of these, we highlight certain objects for some particular information. We highlight D9 since it has been confirmed to be a binary \citep{peisskerBinarySystemCluster2024}. The highlighted X7 is in disagreement in orbital elements calculated by different works for the same object, which is why they are compared to our results separately. The highlighted D2/G3 is a potential binary merger that created X7 \citep{ciurlo_swansong_2023}. The histograms on the sides show in addition the number count averaged over all the simulations. In the histograms we also include the distribution of separated binaries in our simulation. In the top panel, the pink line indicates an increasing binary fraction as radial distance increases, a phenomenon observed in the Galactic Center. The right panel shows the number count of all different objects over eccentricity. Both histograms confidently demonstrate compatibility with the observed objects, aligning seamlessly with our simulations.}
    \label{fig:a_vs_e_all}
\end{figure}

Locations of the merger events as displayed in ~\autoref{fig:merged_binaries_raduis_mass_time} serve as an effective tool to distinguish the effect of the KL oscillations on the inner and outer orbits before merging.
However, the merger products continue orbiting {\sgra} and due to relaxational processes the orbits lose memory of their origin on a time-scale of the order of few $\myr$ and will be observed with different orbital elements and different location with respect to the merger event. We therefore evaluate separately the end states of three types of objects in our simulations which may also be distinguished observationally: (1) surviving primordial binary stars (note that the number of dynamically formed binaries is negligible in our setup), (2) single stars/separated binaries (formerly binaries whose companion was stripped away), and (3) merger products . 

The surviving binaries are located predominantly farther away from the SMBH than the other two groups of objects, with the maximum number count at $a\gtrsim200\,\mpc$ (see the top panel of ~\autoref{fig:a_vs_e_all}). Still some of them have semi-major axes below the original inner edge of the stellar disk due to the disk radial dispersion \citep{subr_two-body_2014} . The presence of single stars on tightly bound orbits is a consequence of the Hills mechanism, which has been suggested as a source of S stars in a similar setup by \citet{subrPropertiesHypervelocityStars2016} along with other works that study the eccentric disk instability and the Hills mechanism as a source of HVS \citep{generozovHillsMechanismGalactic2020b,generozov_forming_2022,akiba_formation_2025}.

The merged objects have a histogram with a maximum close to the initial inner edge of the disk, with a rather flat histogram within the range of $10\,\mpc\lesssim a \lesssim 200\,\mpc$. We find $47.8 \pm 7.0$ merger products ($34.4 \pm 6.2$ for $M_\mathrm{mp}<20 M_\odot$) with $a < 40 \mpc$, that is, within the S cluster. The number of dusty sources currently observed in the S cluster is smaller, which could be due to observational restrictions, but also after a certain period, the merged stars are supposed to settle back on the main sequence and lose the dusty envelope. Some examples of these merger products have been observed in the Galaxy \citep{tylenda_progenitor_2005,tylenda_evolution_2016}. As shown in recent work by \citet{vynatheyaCollisionMergerProducts2026}, merger products are ``puffed up," extended, and fast-rotating. Thus, after the merged stars settle back in the main sequence, they may still be identified among the other young stars that have higher rotation velocities and/or peculiar chemical composition \citep{schneider_stellar_2019, ferraro_fast_2023}. 

An estimate for a dust embedded star of luminosity $L\sim 10 L_\odot$, photosphere radius of $R\sim1 $ au and a mass of $M_\mathrm{mp}\sim 4 M_\odot$ we get a thermal time estimate for the star to contract back to main sequence of $\approx 2.3 \times10^5 $years; however, the lifetime of these dusty objects is up to debate as it can both increase and decrease by various effects, which are discussed more in \autoref{lifetimes}. If we generously approximate the estimated lifetime of a dusty object to be $1\myr$, then we would have $ \sim 4.2 \pm 2.3$ ($\sim 3.3 \pm 2.1$ for $M_\mathrm{mp} \leq 20 M_\odot$)\footnote{This number assumes merger products produced within the last 1 Myr from the simulation end at $\sim 6 \myr$, but the numbers will increase if we end the simulation earlier and look at merger products from there.} dusty sources within the inner S-cluster, while the rest of the merger products, $ \sim 43.6 \pm 6.6$ ($\sim 31.1 \pm 5.8$ for $M_\mathrm{mp} \leq 20 M_\odot$) would have settled back onto the main sequence, out of which a small subset could result in fast rotating stars in the inner S-cluster. 

We compare orbital properties of the merger products to the observed dust-embedded objects in our Galactic Center. Since there are currently only a few dusty sources identified with measured orbital elements, we cannot perform thorough statistical analysis. \citet{Peissker2024a} reports orbital elements of 12 dusty sources, of which 6 have $a_\mathrm{outer} \leq 40 \mpc$, similarly \citet{Ciurlo2020,ciurlo_swansong_2023} list 5 dusty objects, of which 2 have $a_\mathrm{outer} \leq 40 \mpc$. Unfortunately, orbital elements attributed to the sources also suffer from large uncertainties as is shown in ~\autoref{fig:a_vs_e_all}. As an extreme example, orbital solution for the source identified as X7 differs by nearly one order of magnitude in terms of semi-major axis between the recent works of different groups of observers \citep{ciurlo_swansong_2023,Peissker2024a,peisskerClosingGapFollowup2025}. This is due to the debated nature of X7, which could be a dust embedded star \citep{Peissker2024a} or a collection of dust and gas inspiraling into \sgra \citep{ciurlo_swansong_2023}. 
A potentially related dusty object is D2/G3. which is suggested to be a stellar merger that created X7, a collection of gas and dust ejected during the collision process \citep{ciurlo_swansong_2023}. The position of D2/G3 also coincides with the region where we expect merger products.

Somewhat unique among the dusty sources is D9, which has recently been confirmed to be a binary system \citep{peisskerBinarySystemCluster2024}. It is located close to the maximum of the 2D density plot of semi-major axes and eccentricities (main panel of Fig.~\ref{fig:a_vs_e_all}) of the merged objects. At the same time, the region of the $a-e$ space where D9 is located is occupied with surviving binaries with an abundance similar to that of the merged objects. This is compatible with D9 being a binary and as suggested by \citet{peisskerBinarySystemCluster2024} is likely to undergo coalescence in the near future. We note that several other dusty sources are located in the region where, according to our model, surviving binaries and merged objects have similar abundances. This suggests a possibility of them being either recent mergers or surviving binaries. Still, \autoref{fig:a_vs_e_all} shows that the overall shape of the histogram of the merged objects from the model coincides very well with observations of dusty objects, regardless of their origin.


A direct implication of the above described histogram of number counts over semi-major axes is a lower abundance of binaries orbiting \sgra with smaller semi-major axis. \autoref{fig:a_vs_e_all} shows that our model predicts that the binary fraction among young stars originating in a disk in the Galactic Center will grow from zero value at $a \lesssim 20 \mpc$ to more than $50\%$ above $\approx 0.5 \pc$\footnote{Let us note that the maximum value of the binary fraction can change depending on the initial value ($100\%$ in our model), the overall trend should remain unchanged, however.}. This behavior is broadly consistent with observations that report an increasing binary fraction with distance from \sgra in the young nuclear cluster \citep{chuEvidenceDecreasedBinary2023,gautam_estimate_2024}. However, we note that these observational studies are not restricted to binaries formed or present within the stellar disk. The observed population may therefore include additional binaries that formed outside the disk and would contribute to the measured binary fraction. Our results should thus be interpreted as describing the evolution of the disk-born binary population, which likely constitutes only a subset of the observed systems.

The distribution of eccentricities of all three classes of objects is broad, covering the whole range from zero to unity. We find the best fit for eccentricity distribution for each population is a Beta function. The initial disk-born state fits well to Beta($\alpha = 0.68, \beta = 2.31$), which is characterized by a circular-dominated ``cold" profile, where $\alpha < 1$ represents the primordial spike at $e \approx 0$. Following dynamical evolution, all three classes of objects shift toward a ``thermal" distribution ($f(e) = 2e$), which represents the statistical equilibrium limit where Beta($\alpha = 2$ $\beta = 1$). The surviving binaries remain the most ordered or ``sub-thermal" population Beta($\alpha \approx 2.37, \beta \approx 3.16$), maintaining a mode near $e \approx 0.4$. In contrast, single-star populations show advanced signs of dynamical heating; the separated binaries Beta($\alpha \approx 2.06, \beta \approx 2.27$) have reached the thermal $\alpha$-threshold, while merger products are the most randomized ($\alpha \approx 1.85, \beta \approx 1.70$). With the lowest value of $\beta$ among the three groups, the merger products sit closest to the thermal limit, indicating that the collision process effectively erases the initial disk-like orbital memory. Similarly to the case of semi-major axes, we state that eccentricities of the observed dusty sources are in agreement with the properties of merged stars in our model.

\begin{figure}
    \centering
    \includegraphics[width=0.6\linewidth]{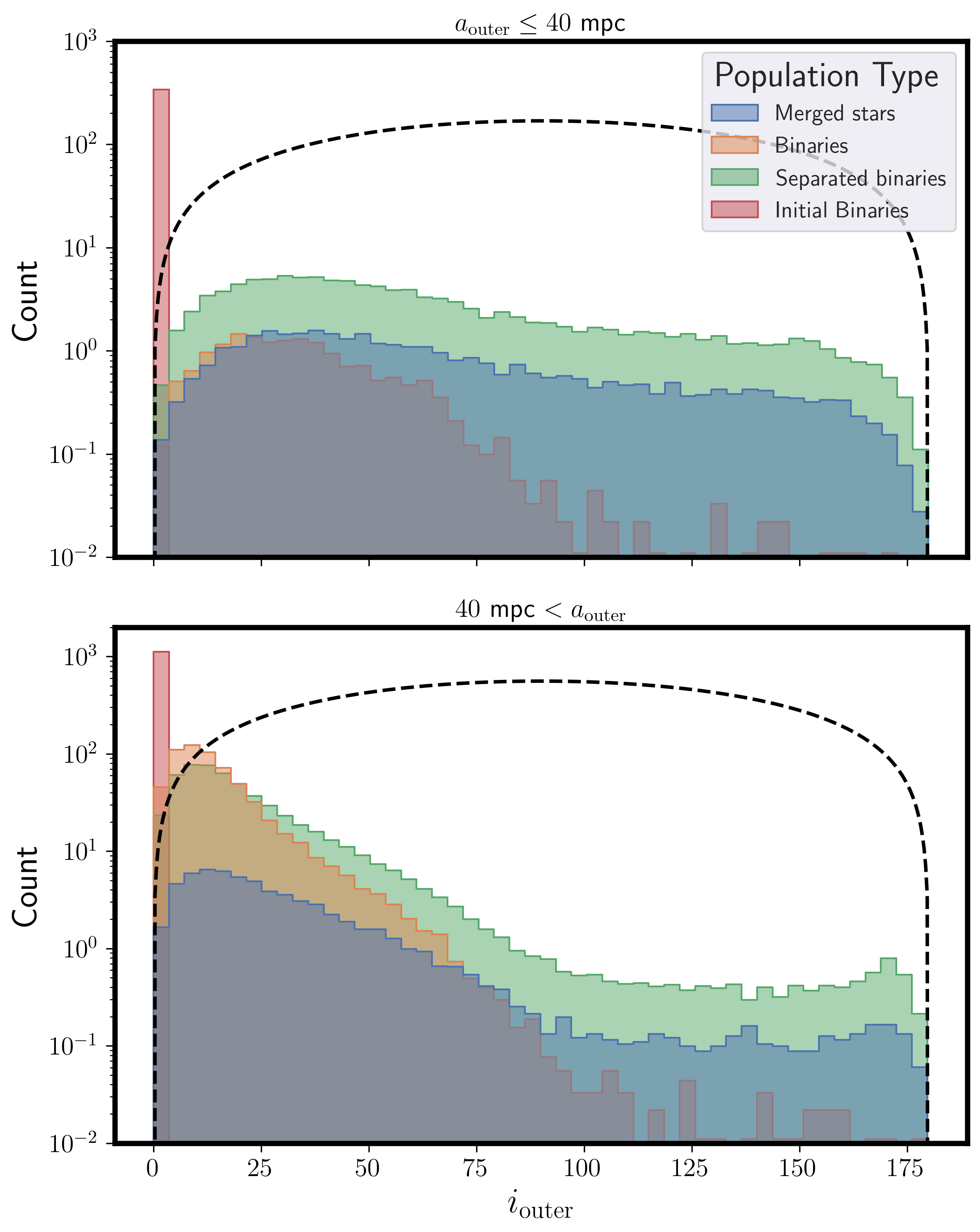}
    \caption{Distribution of outer orbital inclinations for stellar systems after 6 Myr, shown separately for systems with $a_\mathrm{outer} \leq 40$ mpc (top panel) and $a_\mathrm{outer}>40$ mpc (bottom panel). The distributions are divided into initial binaries (red), merged stars (blue), surviving binaries (orange), and separated binaries (green). The black dashed lines shows an isotropic distribution normalized to the number of stars as in the initial binaries. At small radii (top panel), the inclination distribution is significantly broad and approaches isotropic distribution. At larger radii (bottom panel), the distribution becomes significantly broader, but retains a preference for lower inclinations, reflecting partial memory of the initially thin disk configuration}
    \label{fig:inclination_distribution}
\end{figure}

We also compare the inclination distribution of our simulated systems with that of the young stellar population in the Galactic Center. Observations indicate that the innermost region is characterized by a more isotropic distribution of stellar orbits, while at larger radii a coherent clockwise disk is present, albeit with a significant fraction of stars dispersed out of the disk plane \citep{bartkoEvidenceWarpedDisks2009a, luLARPOPULATIONSCENTRAL2013,yelda_properties_2014,vonfellenbergYoungStarsGalactic2022b}. \autoref{fig:inclination_distribution} shows the inclination distribution of the outer orbits after 6 Myr along with the initial binary distribution. In the inner region ($a_{\mathrm{outer}} \leq 40$ mpc), the inclination distribution is significantly broadened and approaches an isotropic configuration (inclinations in that region are supposed to be altered by vector resonant relaxation, as suggested by \citet{hopmanResonantRelaxationMassive2006} and demonstrated by \citet{subr_two-body_2014}.). This is consistent with most observations of the innermost Galactic Center, where the stellar population does not exhibit a strong preference for a common orbital plane.

At larger radii ($a_{\mathrm{outer}} > 40$ mpc), the distribution remains broad but retains a clear excess at low inclinations. This indicates that, while a substantial fraction of systems are dynamically excited out of the initial disk, a non-negligible population still preserves memory of the original thin disk configuration. This behavior qualitatively resembles the observed structure of the clockwise disk, where a coherent disk component coexists with a population of stars on more inclined orbits.

The emergence of an isotropic component in the inner region and a partially coherent disk at larger radii suggests that dynamical evolution alone can reproduce key qualitative features of the Galactic Center stellar distribution, even when starting from an initially thin disk ($i \in [0, 2.5^\circ]$). However, additional sources of perturbations, such as the gaseous circum-nuclear disk, may counter the redistribution of orbital inclinations\citep{martinSurvivingHoleSpatially2012, liuMILKYWAYSUPERMASSIVE2012, hsiehMolecularGasFeeding2017,tsuboiALMAViewCircumnuclear2018,goicoecheaHighspeedMolecularCloudlets2018,hsiehCircumnuclearDiskRevealed2021}. The works of
\cite{haasCouplingYoungStellar2011} and \cite{singhalDynamicalCouplingKeplerian2024}
present the VHS mechanism, which shows that perturbations can preserve disk-like structures closer to the central dominant body, while regions farther away are more affected by relaxational effects. This could possibly explain the observations of \citet{Peissker2024a}, who observe disk-like clustering of the 12 measured dusty objects. However, the magnitude of the effect of the VHS mechanism on our simulations is currently unknown and will be explored in future studies.

Overall, currently our results indicate that an initially thin stellar disk can evolve into a configuration that qualitatively matches the observed Galactic Center structure: an isotropic inner population and a partially coherent outer disk.

\section{Conclusions} 
\label{sec:conclusion}

In this work, we investigated the dynamical evolution of a disk of binary stars orbiting the supermassive black hole \sgra using direct N-body simulations. By modeling the interplay between the disk's self-gravity and the gravitational field of the SMBH, we explored the formation of dusty objects via stellar mergers. Our main conclusions are as follows.

\begin{itemize}
    \item \textbf{Mechanisms of Merger:} The simulations reveal two distinct populations of stellar mergers as shown in \autoref{fig:merged_binaries_raduis_mass_time} and discussed in Section \ref{sec:stellar_mergers} . The first group, which represents the majority of mergers, consists of early mergers (which may be partly unrealistic due to initial conditions) occurring within the main stellar disk due to random mutual perturbations and KL oscillations of the binaries' inner orbits (for example, left panel of \autoref{fig:merger_examples}). The second distinct group occurs within the innermost S-cluster ($d \le 10$~mpc) and is driven by the eccentric KL mechanism, which first excites outer orbital eccentricities and drives binary systems towards \sgra after which the eccentric KL mechanism of the binaries' inner orbits or chaotic three body dynamics involving \sgra (or both) causes coalescence (for example, right panel of \autoref{fig:merger_examples}). 
    
    \item \textbf{Consistency with Observations:} \autoref{fig:a_vs_e_all} shows the orbital elements (semi-major axes and eccentricities) of the simulated merger products, and they exhibit agreement with the population of dust-embedded objects currently observed in the Galactic. We discuss in Section \ref{sec:orb_ele_merger_products} that the model with our initial conditions predicts the existence of approximately $47.8 \pm 7.0$ merger products ($34.4 \pm 6.2$ for $M_\mathrm{mp}<20 M_\odot$) within the inner S-cluster ($a < 40$~mpc) for the duration of the simulation. However, only a subset of those are possibly observed as dusty sources because of their lifetimes, which is not well known. For an estimated lifetime of 1 Myr, we would expect $ \sim 4.2 \pm 2.3$ ($\sim 3.3 \pm 2.1$ for $M_\mathrm{mp} \leq 20 M_\odot$) merger products within the inner S-cluster.

    \item \textbf{Another way of detection:} Some stellar merger products could have settled back on the main sequence and lost their dusty envelope. A fraction of those may be detectable as fast rotators and/or have a specific chemical composition. Based on rough estimates of lifetimes presented in the previous point, we can expect around $ \sim 43.6 \pm 6.6$ ($\sim 31.1 \pm 5.8$ for $M_\mathrm{mp} \leq 20 M_\odot$) stars that could have settled back onto the main sequence within the inner S-cluster ($a<40 \mpc$).
    
    \item \textbf{Binary Fraction Gradient:} The simulations naturally reproduce the observed radial dependence of the binary fraction around Sgr~A*. We see in the top panel of \autoref{fig:a_vs_e_all} that the binary fraction drops to zero in the immediate vicinity of the SMBH ($a \le 20$~mpc) due to the Hills mechanism and mergers, while rising to over 50\% at distances beyond 0.5~pc.

\end{itemize}

These results suggest that stellar mergers induced by secular dynamical evolution are a robust and statistically consistent formation channel for the puzzling dusty objects observed in the Galactic Center. Binary mergers thus play an important role in shaping the stellar content in the Galactic Center.

\begin{acknowledgments}
MS is supported by the Grant Agency of Charles University under the grant number 179123. MZ acknowledges the ﬁnancial support of the Czech Science Foundation Junior Star grant no. GM24-10599M. LS, JH and MS acknowledge the support of Czech Science Foundation grant no. 26-21774S. FP gratefully acknowledge the Collaborative Research Center 1601 funded by the Deutsche Forschungsgemeinschaft (DFG, German Research Foundation) – SFB 1601 [sub-project A3] – 500700252.
\end{acknowledgments}

\begin{contribution}

M.S. ran all the simulations, did the analysis, and led the writing of the manuscript. L.S. and J.H. contributed codes and tools to perform the simulations along with discussions to analyze the results. M.Z. contributed for the discussion of lifetime of dust-embedded objects. F.P. contributed to the interpretation of observations of dusty objects.  All authors contributed to the writing of the manuscript.


\end{contribution}

%

\software{NBODY6 \citep{Aarseth_2003},  
          Astropy \citep{astropy:2013, astropy:2018, astropy:2022}, 
          NumPy \citep{harris2020array},
          Matplotlib \citep{Hunter:2007},
          ARWV \citep{chassonneryARWVCodeUser2019},
          REBOUND \citep{reinREBOUNDOpensourceMultipurpose2012},
          REBOUNDx \citep{tamayoREBOUNDxLibraryAdding2020}
          }


\appendix

\section{Impact of initial conditions} 
\label{app:ini_impact}

\autoref{tab:model_ini} summarizes the initial conditions adopted for the model investigated in this study (M1) together with two additional ones with different parameters (M2 \& M3). The models differ in their choices of minimum and maximum outer semi-major axes ($a_\mathrm{min}$, $a_\mathrm{max}$) and the ranges for the inner binary semi-major axes ($a_\mathrm{inner,low}$, $a_\mathrm{inner,high}$). Model M1, which is the primary focus in the main text, utilizes the smallest $a_\mathrm{min}$ value, while M2 and M3 employ moderately larger values. Notably, M2 allows for tighter inner binaries compared to the other models. These differences are intended to test the sensitivity of the binary evolution outcomes to variations in the initial configuration. Let us also note that the model M3 has the same initial conditions as in \citet{subrPropertiesHypervelocityStars2016}.

\begin{figure}
    \centering
    \includegraphics[width=0.5\linewidth]{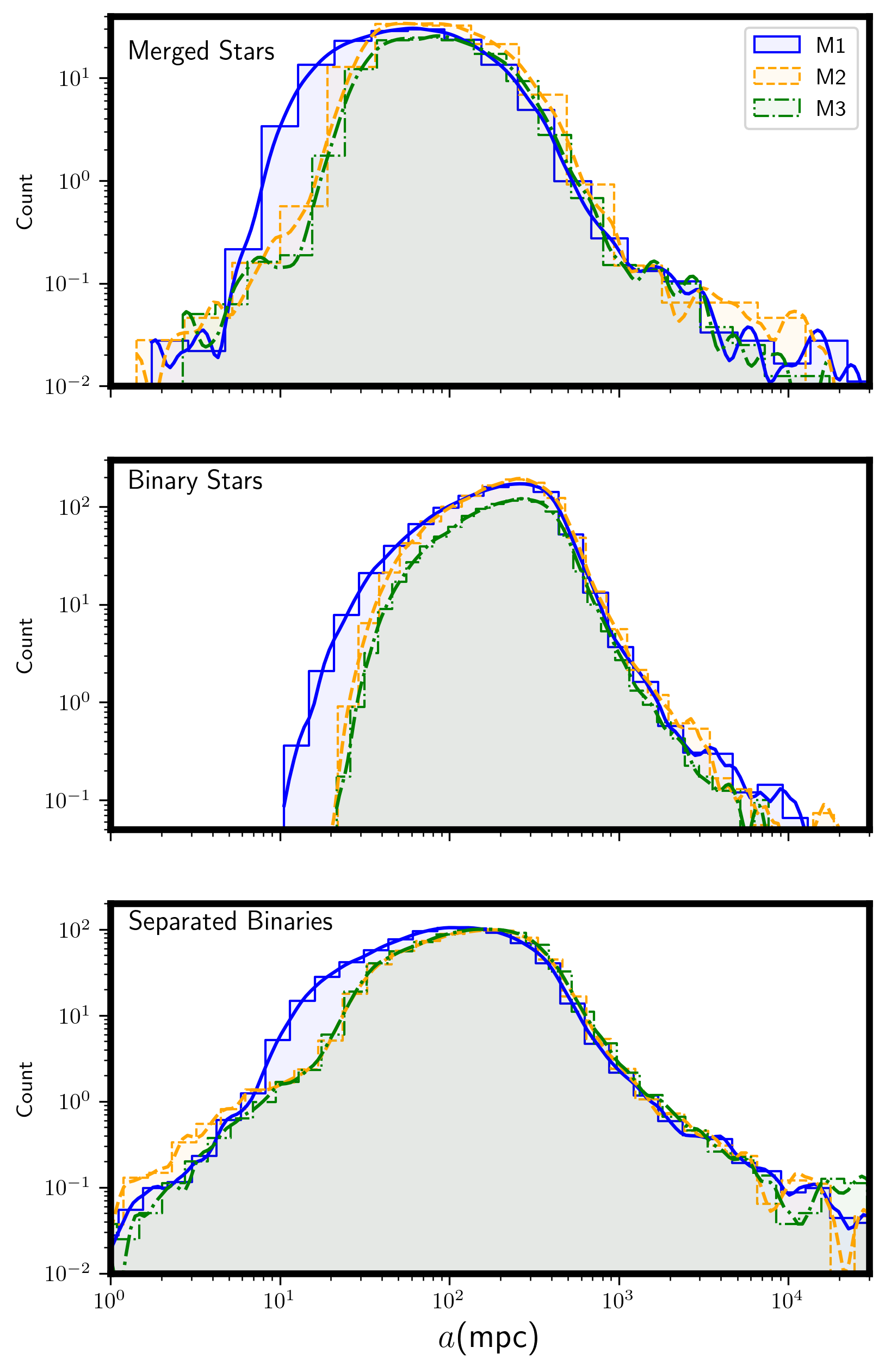}
    \caption{Impact of initial conditions on the final semi-major axis. In the histograms, we show the final semi-major axis histogram of merged stars, binary stars, and separated binaries (from top to bottom) for three different models with varying initial conditions as described in \autoref{tab:model_ini}. The merged stars and binary stars show that for M1, the lower value of $a_\mathrm{min}$ results in a slight increase in the counts of these objects at lower semi-major axes. The distributions for M2 and M3 suggest that the final states of merged and binary stars are largely insensitive to the specific initial distribution of $a_\mathrm{inner}$ within the tested range. However, in the last panel, the smaller $a_\mathrm{inner}$ of M2 results in tighter orbits around \sgra. Distribution of separated binaries in M1 shows a bump near $a_\mathrm{min}$, reflecting the initial setup. Overall, while the initial conditions exert a measurable influence on the specific orbital parameters, the qualitative evolutionary outcomes of the system remain consistent across the models. }
    \label{fig:ini_cond_comp}
\end{figure}

\autoref{fig:ini_cond_comp} presents a comparison of the final spread of semi-major axes for merged stars, binary stars, and separated binaries across the three models at 6 Myr. The histograms indicate that a lower value of  $a_\mathrm{min}$ in M1 results in a modest excess of objects with smaller semi-major axes. For separated binaries, M1 exhibits a minor bump near $a_\mathrm{min}$, reflecting the imposed initial boundary. In contrast, the histograms for M2 and M3 show that the range of $a_\mathrm{inner}$ has a minimal effect on the populations of merged and binary stars. The lower limit of $a_\mathrm{inner}$ in M2 results in a slight increase in the number of separated binaries with $a < 10$ mpc. Overall, these comparisons demonstrate that while initial conditions introduce some variation in the final histogram, the general evolutionary trends and population outcomes remain consistent among the models.

\begin{table}[]
    \centering
    \begin{tabular}{c|c|c|c|c}
    Name & $a_\mathrm{min}$ & $a_\mathrm{max}$ & $a_\mathrm{inner,min}$ & $a_\mathrm{inner,max}$\\
    \hline
         M1&0.02pc&0.4pc&0.1au&100au  \\
         M2&0.04pc&0.4pc&0.05au&100au\\
         M3&0.04pc&0.4pc&0.1au&100au
    \end{tabular}
    \caption{Initial conditions of different models, with M1 being the model we presented in the main text. M2 and M3 have a higher value of $a_\mathrm{min}$ with M2 also having tighter binaries.  }
    \label{tab:model_ini}
\end{table}

\section{Limitations of simulations}
\label{app:limitations}

The simulations use various simplifications to allow us to have a large statistical sample for the analysis. One main limitation is that our simulations do not include stellar evolution and mass loss. This mass loss and evolution would be crucial for massive stars, but in this work, we are primarily interested in lower-mass stellar mergers, which spend a much longer time on the main sequence and have lower mass-loss rates.

Additionally, our simulations have a 100\% efficient merger process, which is unrealistic. For instance, \citet{roseModelingStellarCollisions2026} used a suite of SPH simulations to develop a framework for predicting mass loss and trajectory changes in stellar collisions within galactic nuclei. Similarly, \citet{vynatheyaCollisionMergerProducts2026} utilized the AREPO code to demonstrate that while both collisions and mergers produce massive, rapidly rotating stars, they differ significantly in their core hydrogen fractions and magnetic field structures. Due to the high computational cost and the different physical regimes involved, the detailed outcomes of these hydrodynamic works cannot currently be implemented directly into our N-body model. Integrating such realistic merger prescriptions—including non-isotropic mass loss and post-collision velocity kicks—remains a complex challenge that should be explored in future work. The final results should not vary much since we are comparing the orbital elements of the final merger products to the observed dusty objects, which, if the mass loss is minimal and isotropic, would not change substantially.

We also set up our simulations to have all stars on the main sequence at $T=0$. However, star formation does not happen all at once, especially for lower-mass stars, which can result in a delayed formation of lower-mass binaries that are the progenitors of the dusty objects. This could delay the mergers and spread out the merger peaks at $T=0$ and $T\approx1.5$ Myrs. The same effect would be observed in case the disk is enriched by an inflow of more material, leading to another, more recent star formation event.



\section{Impact of GR}
\label{sec:gr_consequences}

In our setup, relativistic apsidal precession induced by \sgra can affect both the orbit of the binary center of mass around \sgra and the internal orbit of the binary itself, particularly when the binary system approaches small pericenter distances in its orbit around \sgra. KL oscillations of both the inner and outer orbit may be suppressed when relativistic precession becomes sufficiently rapid. Therefore, we explore the impact of relativistic precession on both the outer orbit of the binary center of mass and the inner orbit of the binary during close passages to \sgra. We find that while relativistic precession can suppress KL oscillations in parts of parameter space, it does not fully eliminate them in more realistic disk environments.

\subsection{Outer Orbit}
\label{sec:outerorbit}

Relativistic precession induced by \sgra can affect our simulation results, as outer-KL oscillations may be suppressed for systems with small semi-major axes and high eccentricities. To estimate which systems are most affected, we compare the relativistic apsidal precession rate with the characteristic timescale of KL oscillations of the outer orbit. KL oscillations require a slow and coherent evolution of the argument of periapsis, whereas relativistic effects introduce rapid precession that can break this resonance.

The relativistic precession rate is given by
\begin{equation}
\dot{\omega}_\mathrm{GR} = \frac{3GM_\bullet}{ac^2(1-e^2)}\sqrt{\frac{GM_\bullet}{a^3}}
\label{omegadot_gr}
\end{equation}
where $M_\bullet$ is the mass of \sgra, $a$ and $e$ are the semi-major axis and eccentricity of the orbit, and $c$ is the speed of light. This expression shows that relativistic precession becomes stronger for smaller semi-major axes and higher eccentricities.

We approximate the KL precession rate as
\begin{equation}
\dot{\omega}_{KL}\sim 1/T_\mathrm{KL}
\end{equation}
where $T_\mathrm{KL}$ is the characteristic KL timescale. When
$\dot{\omega}_\mathrm{GR} \gtrsim \dot{\omega}_\mathrm{KL}$
relativistic precession dominates, and KL oscillations are expected to be suppressed.

\begin{figure}
    \centering
    \includegraphics[width=\linewidth]{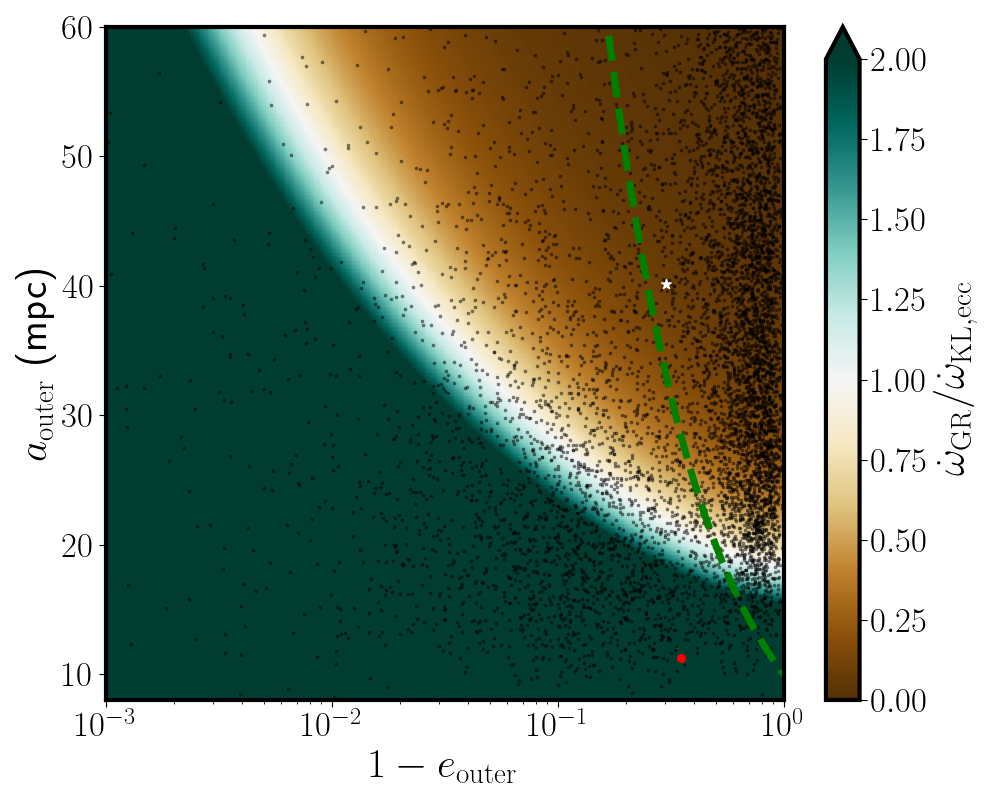}
    \caption{In the heatmap we show the ratio of $\dot{\omega}_\mathrm{GR} / \dot{\omega}_{KL}$ for the outer orbit of the binary system around \sgra at $a_\mathrm{outer}$ with an eccentricity of $e_\mathrm{outer}$. The perturber is a single particle mimicing the center of mass of the disk $M_\mathrm{pert}=24500 M_\odot$ orbitting at $a_{pert}=0.12\pc$ with an eccentricity of $e_{pert}=0.44$. The region where $\dot{\omega}_\mathrm{GR} / \dot{\omega}_{KL} \gtrsim 1$ is where we expect KL oscillations to be completely damped while for the rest of the parameter space we expect KL oscillations to be undamped. Black dots mark the orbital parameters of merging binaries with $M_c \leq 20 M_\odot$. Red dot marks the orbital parameter of a star that undegoes KL oscillations with GR in a numerical simulation as showed in \autoref{fig:rebound_disk_sim}. The green dashed line represents where $r_\mathrm{p,outer}=10 \mpc$. The white star marks the starting location of the binary shown in \autoref{fig:gr_kl_inner}. } 

    \label{fig:kl_gr_omega_comp}
\end{figure}

In this estimate, we approximate the stellar disk as a single effective perturber located at the center of mass of the disk i.e., we consider a perturber of mass $M_\mathrm{pert}=24500M_\odot$ and orbital parameters $a_\mathrm{pert}=0.12\mathrm{pc}$ and $e_\mathrm{pert}=0.44$. In \autoref{fig:kl_gr_omega_comp} we show the ratio of 
$\dot{\omega}_\mathrm{GR} / \dot{\omega}_{KL}$ for a parameter space of $a_\mathrm{outer}$ and $e_\mathrm{outer}$. Regions where $\dot{\omega}_\mathrm{GR}/\dot{\omega}_\mathrm{KL} \gtrsim 1$ correspond to parameter combinations where relativistic precession occurs on a time scale comparable to or shorter than KL oscillations. In this regime, outer-KL oscillations are expected to be strongly suppressed. The black dots in \autoref{fig:kl_gr_omega_comp} show the orbital elements of binary systems with $M_c \leq 20 M_\odot$ around \sgra at the time of merging, and the green dashed line marks $r_\mathrm{p,outer}=10\mpc$. As we see, a small fraction of the merging binaries are in the region where our approximation suggests outer-KL oscillations should be damped, and for mergers that happen with orbits with a pericenter of less than 10 mpc, it is more evenly spread between the region where we expect damping of KL oscillations and where we do not.  
However, this approximation neglects the distributed nature of the disk potential and therefore provides only an order-of-magnitude estimate.

\begin{figure}
    \centering
    \includegraphics[width=\linewidth]{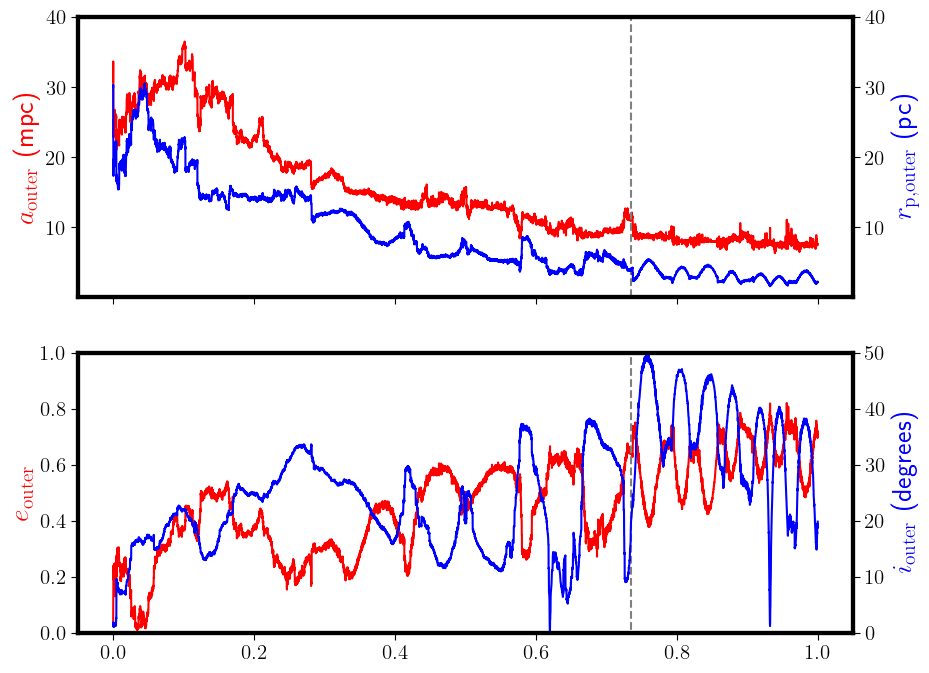}
    \caption{Time evolution of the orbital elements of a representative star in a simplified disk simulation including relativistic corrections using REBOUNDX. The disk consists of 490 stars of $50M_\odot$ each. After an initial phase dominated by local stellar interactions, the orbit exhibits KL-like oscillations despite lying in a region where analytic estimates predict suppression, highlighting the limitations of the single-perturber approximation. The dashed line is represents the time we use for the marker in \autoref{fig:kl_gr_omega_comp}.}
    \label{fig:rebound_disk_sim}
\end{figure}

To test the validity of this approximation, we perform numerical experiments that include relativistic effects in a more realistic stellar disk configuration. Without running accurate simulations of the entire disk of 2000 stars in a relativistic setup, we can study smaller setups with added relativistic effects to estimate effects of relativistic precession on outer-KL oscillations. To do so, we use REBOUND which is a Nbody integrator with the additional REBOUNDX addon which allows us to implement an additional force that mimics the relativistic effects on our orbits (we use the ``gr\_potential" as it was computationally stable)\citep{reinREBOUNDOpensourceMultipurpose2012,tamayoREBOUNDxLibraryAdding2020}. We simulate a disk of 490 stars of mass $50 M_\odot$ each, giving a total mass of the disk of 24500 $M_\odot$, same as the original setup. These stars were placed in the same orbits as the center of masses of the binaries in the original setup. In \autoref{fig:rebound_disk_sim} we see the evolution of a single body in the simplified setup with 50 $M_\odot$ bodies. The evolution initially reflects strong interactions with nearby disk stars. Later, the orbit settled into outer-KL-like oscillations. 

The orbital elements of this body right before it started undergoing outer-KL oscillations is shown by the red dot in \autoref{fig:kl_gr_omega_comp}, demonstrating that even though the approximation suggests the outer-KL oscillations should be damped, they are present and thus for a selection of mergers in our simulation, we can still expect outer-KL oscillations to be present in their outer orbits.

\subsection{Inner Orbit}
\label{sec:inner_orbit_gr}
The inner orbit of the binary system also undergoes inner-KL oscillations with \sgra as the perturber. We perform some 4 body integrations of a binary in orbit around \sgra with a distant perturber mimicking the disk of stars. As discussed previously, the single particle approximation is not completely correct but it does allow an estimate on what undergoes at those scales. We use ARWV, a few body integrator which uses the AR-Chain algorithm to integrate a few bodies accurately with upto 2.5 order post-Newtonian corrections
\citep{chassonneryARWVCodeUser2019}.

The damping due to relativity could affect the inner-KL oscillations for tight binaries, and can be calculated with the orbital elements of the inner binary. We have performed an analysis of $\dot{\omega}_\mathrm{GR}/\dot{\omega}_\mathrm{KL}$ for a binary system of $m_1=1.5 M_\odot$ and $m_2=1 M_\odot$ orbiting each other at a separation of $a_\mathrm{inner}=0.1$ AU, with an initial eccentricity of $e_\mathrm{inner}=0.01$ and an initial inclination of $i_\mathrm{inner}=0^\circ$ (with respect to the plane of the outer orbit). The perturber is modeled as a single massive particle with the same as the center of mass of the disk as described in the Appendix ~\ref{sec:outerorbit}

\begin{figure}
    \centering
\includegraphics[width=\linewidth]{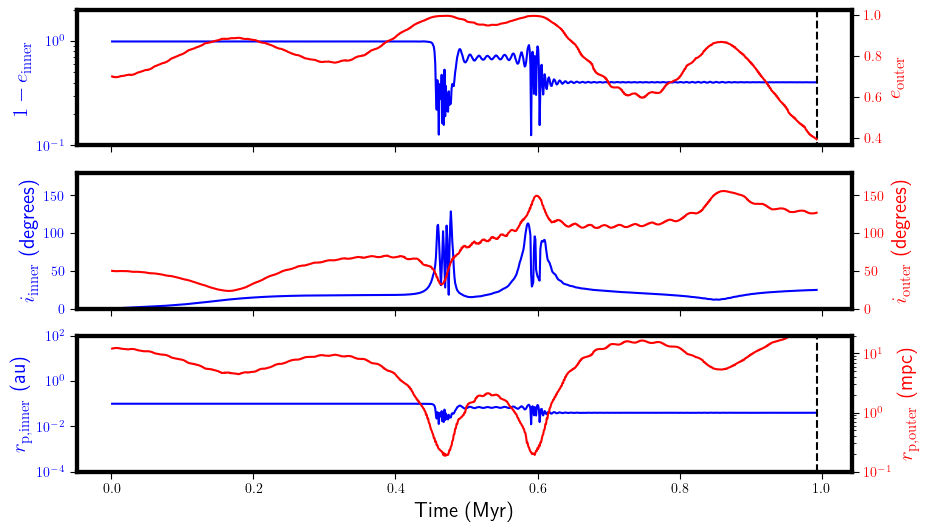}
    \caption{Evolution of the inner and outer orbital elements of a binary system orbiting Sgr A* with initial outer inclination $i_\mathrm{outer}=50^\circ$, including post-Newtonian corrections, with an initial semi-major axis $a_\mathrm{inner}=0.1$ AU and outer inclination of $i_\mathrm{outer}=50^\circ$. The binary system is placed on an eccentric orbit around \sgra with $a_\mathrm{outer}=40\ \mpc$ and $e_\mathrm{outer}=0.7$. Strong inner eccentricity oscillations are evident when the outer orbit becomes more eccentric, and its pericenter decreases, leading to inner eccentricities that could result in a merger before disruption in a more complete physical model.}
    \label{fig:inner_KL_gr_tighter}
\end{figure}

\autoref{fig:gr_kl_inner} shows the heatmap of $\dot{\omega}_\mathrm{GR}/\dot{\omega}_\mathrm{KL}$ for the binary, with the value of $\dot{\omega}_\mathrm{GR}/\dot{\omega}_\mathrm{KL}\gtrsim 1$ are the regions where the inner-KL oscillations are damped due to GR according to the analytical approximations. However, as we showed earlier, along with in  \autoref{sec:outerorbit}, the outer orbit undergoes outer-KL oscillations which change where the binary is located in this plot. The red dotted line in \autoref{fig:gr_kl_inner} shows the orbital evolution of a binary with $a_\mathrm{inner}=0.1$ AU, $a_\mathrm{outer}=40\ \mpc$ and $e_\mathrm{outer}=0.7$. Its initial condition is shown by the white star. The binary starts in the region where the inner-KL oscillations should be damped; however, as it undergoes outer-KL oscillations, it passes through regions where we can expect inner-KL oscillations. The evolution of the orbit is shown in detail in \autoref{fig:inner_KL_gr_tighter}. In the top panel, we see the eccentricity of both the inner and outer binary, in the second panel, we see the inclination of the inner and outer binary around \sgra, and in the third panel we see the pericenter distances of the inner and outer binary. We can see that when the outer orbit reaches high eccentricity due to outer-KL oscillations, the inner orbit undergoes rapid eccentricity oscillations, which could result in close flybys and mergers.    

\begin{figure}
    \centering
    \includegraphics[width=\linewidth]{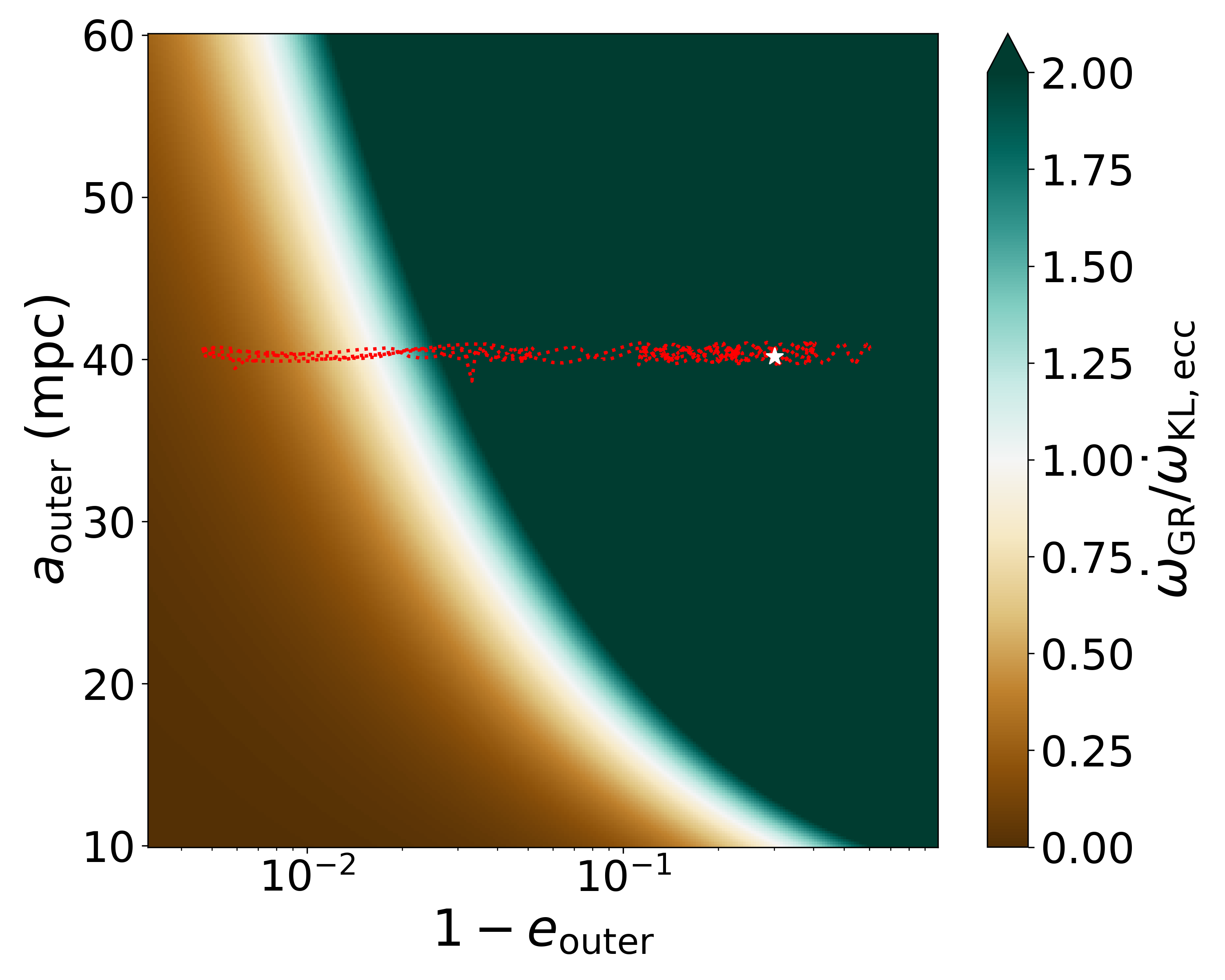}
    \caption{A heatmap showing the ratio of $\dot{\omega}_\mathrm{GR}/\dot{\omega}_\mathrm{KL}$ for the inner binary system orbiting \sgra for the same setup as shown in \autoref{fig:kl_gr_omega_comp}, with the inner binary having $a_\mathrm{inner}=0.1 $ AU. Similar to \autoref{fig:kl_gr_omega_comp}, $\dot{\omega}_\mathrm{GR}/\dot{\omega}_\mathrm{KL} \gtrsim 1$ is where we expect KL oscillations of the inner binary to be completely damped. However due to outer KL oscillations, we expect the binary to migrate from region with damped KL oscillations to undamped KL oscillations as shown by an example evolution of a binary system which started at the location shown by the white star. The evolution of this system in $a-e$ space is marked by the red dotted line, and shown in detail in \autoref{fig:inner_KL_gr_tighter}.  }
    \label{fig:gr_kl_inner}
\end{figure}

Close fly-bys of other stars can also affect the orbit of the binary stars, as seen for example in \autoref{fig:merger_examples}, which can cause the inner orbit to change, resulting in a binary that could not undergo inner-KL to start undergoing KL oscillations. Close fly-bys like this could also affect binaries affected by other short-range forces, such as tidal circularization, changing their orbital elements such that inner-KL oscillations are no longer suppressed by relativistic precession. \citep{winter-granicBinaryMergersCenters2024}

Overall, relativistic corrections can modify both inner and outer KL dynamics, modestly reduce merger and ejection rates, but mergers driven by these secular and chaotic dynamics remain a robust prediction due to the active orbital evolution driven by the surrounding disk and star encounters.

\section{Merger Lifetimes} 
\label{lifetimes}
We suggest that the stellar mergers that are produced in our numerical model are counterparts of the dusty sources observed in the Galactic Center. The match between the model and observations, however, does not need to be accurate. One important source of discrepancy can be caused by not considering internal evolution of stars, including the mergers, in the model.
The merger products discussed in \autoref{sec:results} were analyzed over the entire duration of the simulation. However, the stellar objects that form via the collision of binary components do evolve in time. Hydrodynamic simulations showed that stellar merger remnants are initially far from hydrostatic and thermal equilibrium, often possessing inflated envelopes that also contain dust beyond the sublimation radius. The excess internal and mechanical energy deposited during the collision is subsequently radiated away as the remnant contracts on the Kelvin–Helmholtz (thermal) timescale \citep{1990sse..book.....K,Suzuki2007,2009A&A...497..255G,2023ApJ...955...30R,2025arXiv250918421S}. On that time-scale, we assume that its dusty envelope tends to disappear. Hence, only the recent mergers should be interpreted as dusty sources. Considering the typical properties of dust-embedded stars with the luminosity of $L_{\rm c}\sim 10\,L_{\odot}$ and the photosphere radius of $R_{\rm c}\sim 1\,{\rm AU}$ \citep{Peissker2020b} and taking into account the mean merger product mass of $M_{\mathrm{mp}}\sim 4\,M_{\odot}$ (S cluster region; see ~\autoref{fig:merged_binaries_raduis_mass_time}), the typical thermal timescale for the whole volume of the dusty object can be estimated as,
\begin{align}
t_{\rm th} &\sim \frac{GM_{\mathrm{mp}}^2}{R_{\rm c} L_{\rm c}}\,\notag\\
           &\sim 2.3\times 10^5 \left(\frac{M_{\mathrm{mp}}}{4\,M_{\odot}}\right)^2 \left(\frac{R_{\rm c}}{1\,\rm{AU}}\right)^{-1} \left(\frac{L_{\rm c}}{10\,L_{\odot}}\right)^{-1} \text{years}. 
\label{eq:KH_timescale}
\end{align}

In our simulations, mergers are completely efficient, meaning that all the mass of the merging stars is retained in the resulting merger product. In this regard, Eq.~\eqref{eq:KH_timescale} rather provides an upper limit for the relaxation (thermal) timescale (for a given size and luminosity of the merger product) since stellar collisions typically lead to mass loss, which decreases the remnant mass and hence also the estimated Kelvin-Helmholtz timescale. This is in accordance with the fact that the observed dusty sources have a mass typically lower than the typical young stars in the Galactic Center \citep{habibi_twelve_2017,peisskerS6299Yr2020,Peissker2024a,peisskerBinarySystemCluster2024,peisskerClosingGapFollowup2025}.

However, other factors can further affect the dusty envelope lifetime, both decreasing it and prolonging it, specifically the detailed dynamics of the outflows as well as accretion onto the merger product. Contributing effects include tidal stripping of the dusty envelope by the SMBH on the orbital (tidal) timescale \citep{zajacek_dust-enshrouded_2014}, photoevaporation by the ambient UV field, dust-driven wind due to radiation from the embedded star, and the ram pressure stripping due to mildly supersonic motion through the circumnuclear medium \citep{zajacek2016,zajacek_nature_2017}.

The observation of D9, a dusty object confirmed to be a binary system, further suggests the possibility of formation of the dusty envelope during close, yet non-destructive encounters between the binary components. This may prolong the period of the appearance of the object as a dusty source with infrared excess for a long period of time. Recent hydrodynamic modeling by \citet{badouxKozaidrivenMassLoss2026} suggests that a circumbinary disk around D9 can remain stable for several million years with periodic outflows with repeated close encounters during the inner KL oscillations.


\bibliography{sample701}{}
\bibliographystyle{aasjournalv7}



\end{document}